# A quantum uncertainty entails entangled linguistic sequences

F.T.Arecchi[1,2]*

1 University of Florence, Department of Physics, Via Sansone, 1 - 50019 Sesto Fiorentino, Firenze, Italia.

2 INO-CNR, Largo E. Fermi, 6 - 50125 Firenze, Italia.

*e-mail: tito.arecchi@ino.it

## Abstract

Synchronization of finite spike sequences is the way two brain regions compare their content and extract the most suitable sequence. This is the core of the linguistic comparison between a word and a previous one retrieved by memory. Classifying the information content of neural spike trains, an uncertainty relation emerges between the bit size of a word and its duration. This uncertainty affects the task of synchronizing spike trains of different duration representing different words, entailing the occurrence of entangled sequences, so that word comparison amounts to a measurement based quantum computation. Entanglement explains the inverse Bayes inference that connects different words in a linguistic search. The behaviour here discussed provides an explanation for other reported evidences of quantum effects in human cognitive processes lacking a plausible framework, since either no assignment of an appropriate quantum constant had been associated, or speculating on microscopic processes dependent on Planck's constant resulted in unrealistic de-coherence times.

## Main

A finite neuronal spike sequence consists of a sequence of temporal bins, each one labelled as *1* (spike present) or *0* (spike absent). In human language operations, comparison of a word with a previous one recalled by the short term memory entails the synchronization of finite neuronal spike sequences (SFSS) [1,2]. The language operation can be seen as an instance of the hotly debated topics of consciousness retrieval [3]

As a computational tool of brain neuron networks, an uncertainty relation and an associated de-coherence time in SFSS had already been introduced and discussed in detail [4]. Information – time uncertainty (***ITU***) resembling the quantum energy-time uncertainty is here outlined. ***ITU*** is expressed as spike number-time uncertainty. Because of ***ITU***, the comparison of two words, coded as sequences of spikes, entails entangled spike sequences (*ES*). This paper completes the previous one [4] by exploring in detail the role of entanglement in linguistic tasks.

*ES* results from ***ITU***; it provides a computational speed up in the comparison of a word with a previous one recalled by the short term memory (meaning problem), thus connecting two pieces of a linguistic sequence. Such a comparison has been shown as corresponding to inverting the Bayes inference formula and hence called *"inverse Bayes"* [3]. In dealing with a linguistic task, the meaning problem consists of finding the most appropriate match (congruence) between a word just occurred in the last reading of a linguistic piece and an associated word occurring in the previous piece recalled via the short term memory.

The above process is confined within a *de-coherence time*, that in the brain case results to be of the order of a few hundred milliseconds; it is the crucial time allotted for dealing with the meaning problem.

In virtue of the theta-gamma modulation of the EEG [*Supplementary material*], the spike train coding the first word is interrupted from a duration $T$ to a duration $\Delta T < T$. To perform the synchronization, it must be lengthened by $T-\Delta T$ and this can occur in $N=2^{(T-\Delta T)}$ ways by filling the $T-\Delta T$ interval with $N$ different sequences of *0* and *1*. Thus the first word is coded as a cluster state $|E>$ consisting of $N$ different sequences. We show that the state $|E>$ is entangled, namely, that the $N$ component states have quantum correlations.

If e.g. $\Delta T = T-1$, then the synchronization task of $|T>$ with $|\Delta T>$ amounts to comparing $|T>$ with the entangled state

$$|E> = 1/\sqrt{2} \, ( \, |\Delta T, 0> + |\Delta T, 1> \, ) \qquad (1)$$

Without further treatment, the most synchronized state will be identical to the second world. If however the $N$ states are each weighted differently by a semantic operator that we call $\epsilon$ ($\epsilon$ stays for *emotion*), then the most synchronized word emerging is the joint result of $\epsilon$ and of the code of the second world.

Let us define a processing time $T$ as the time it takes for the brain to build up a complete decision like naming a picture or reading a word aloud. It corresponds to the reaction time for visual lexical decisions or word naming, it occurs, in a range from 300 to 700 ms [5]. Then, the total number of binary words that can be processed is $P_M = 2^{T/\tau}$. If e.g. T=300 ms, it follows $P_M \approx 10^{33}$.

Considering the synchronization task between spike trains, we show that interruption of a spike train introduces a joint uncertainty in the word information and spike duration. Let us investigate what brain mechanisms rule the duration time. The threshold for spike onset is modulated by a EEG gamma frequency oscillation around 50 Hz; spike threshold being lowest close to the maxima of the gamma wave. Phase coherence of the gamma wave over distant brain regions permits spike synchronization overcoming delays due to the finite propagation speed in the neural axons [6, 7]. Furthermore, a lower frequency EEG oscillation (theta band, around 7 Hz) controls the number of gamma maxima within a processing interval. The theta- gamma cross-modulation corresponds to stopping the neural sequence at $\Delta T \leq T$ [8, 9]. As a result, all spike trains equal up to $\Delta T$, but different by at least one spike in the interval $T-\Delta T$, provide an uncertainty cloud $\Delta P$ such that [4]

$$\Delta P = 2^{(T-\Delta T)/\tau} = P_M 2^{-\Delta T/\tau} \qquad (2)$$

Thus we have a peculiar uncertainty of exponential type between spike information P and duration T, that is,

$$\Delta P \cdot 2^{\Delta T/\tau} = P_M \qquad (3)$$

Where $P_M = 2^{T/\tau}$. The whole $T$ train is a vector of a $2^T$ dimensional Hilbert space; synchronization of two different $T$ trains amounts to counting the number of 0 and 1 coincidences.

Since synchronization entails equal lengths of the trains under comparison, a $\Delta T$ pulse acquires a length $T$ by being **entangled** with all possible sequences of *0*'s and *1*'s that can be contained in the complementary interval $T-\Delta T$. As we project state |T> over | E>, we extract the congruence as the best synchronization S, namely

$$S = <E|T> \qquad (0<S>1) \qquad (4)$$

For sake of providing an example, let us take $T=10$, $\Delta T=9$; hence, the following synchronization values result: $S_1=(9+1)/10=1$; $S_2=(9+0)/10=0.9$.

Define a fractional bit number $u=P/P_M$; then the fractional uncertainty $\Delta u=\Delta P/P_M$ is related to the gated time $\Delta T$ by

$$\Delta u \cdot 2^{\Delta T/\tau} = 1 \qquad (5)$$

The two conjugated quantities (i.e., fractional bit number $u$ and duration $\Delta T$ of the gated spike train) are coupled by an exponential type uncertainty relation. By a change of variable

$$y = \tau 2^{t/\tau} \qquad (6)$$

where now $y$ has time dimensions, we arrive to a product type uncertainty relation

$$\Delta u \Delta y = \tau \qquad (7)$$

Thus, in the space $(u, y)$ we have a Heisenberg-like uncertainty relation.

In the synchronization of interrupted spike trains, the associated uncertainty constant $C$ can be assigned in $J \times s$, once we explicit the energy per spike. This energy corresponds to the opening along the axon of about $10^7$ ionic channels [10] each one entailing an ATP $\rightarrow$ ADP+P chemical reaction yielding 0.3 eV, thus the minimal energy disturbance in neural spike dynamics is $3 \cdot 10^{-13} J$, that is, around $10^8 k_B T_r$ (where $k_B$ is the Boltzmann constant and $T_r$ is the room temperature). Since $\tau = 3$ *ms*, it results $C \approx 10^{-15} Js \approx 10^{19} \hbar$.

However, due to the structure of Eq. (3), the uncertainty holds over a finite range, between two extremes, namely (measuring times in $\tau$ units)

i) $\Delta T_{min}=1$, yielding $\Delta P_{max}=P_M/2$

and

ii) $\Delta T_{max}=T$, yielding $\Delta P_{min}=1$.

Following the standard procedure of a quantum approach, we expect single sequence interference and two sequence entanglement within such a time interval.

Once an entangled state has been prepared, it lasts over a time that we call de-coherence time.

For $\Delta P=1$ (minimal disturbance represented by 1 spike) we have the de-coherence time

$$\Delta y_d = P_M \tau \qquad (8)$$

Using the numbers already reported above, the de-coherence time scales as $\log_2 P_M=100 \cdot (bins)$, and going from bins to sec: $\tau_d=decoherencetime=0.3 sec$, very far from the value $\tau_d=\hbar/k_B T_r \approx 10^{-14}$s., evaluated for single Newtonian particles disturbed by the thermal energy $k_B T_r$ [11,12].

Notice that the resulting de-coherence time is equal to the full processing time $T=300ms$ chosen as an example. If we consider a different full processing time for the SFSS, the de-coherence time changes accordingly.

The equivalent of the De Broglie wave results as follows. Comparing Eq.(7)) with the Fourier relation $\Delta\omega \, \Delta t \geq 1$, we introduce a wave-like assumption

$$\omega = \frac{1}{\tau}\frac{P}{P_M} = \frac{u}{\tau} \qquad (9)$$

If $N$ is the total spike number over time $T$ and $N'$ the spike number in the interrupted interval $\Delta T$ then

$$u = \frac{P}{P_M} = \frac{2^{N'}}{2^N} = 2^{-(N-N')} = 2^{-(T-\Delta T)/\tau} \qquad (10)$$

The equivalent of the two slit self-interference of a single particle would be the comparison of a single spike train of $P$ bits with a delayed version of itself. A laboratory implementation would consist of the spike train translated in time and superposed to the original train. As one changes the time separation $\Delta t$ of the two trains, the spike synchronization (number of coincidences of *1*'s) decays as soon as $\Delta t>\tau$ from $N$ (number of *1*'s in the spike train) to $\sqrt{N}$ (random overlaps). However, further increasing the time separation, self-interference entails a revival of the synchronization depending on the Fourier periodicity, that is, for

$$\Delta t = \tau \frac{P_M}{P} = \tau 2^{(T-\Delta T)/\tau} = \tau 2^x \qquad (11)$$

where we call $x=(T-\Delta T)/\tau$ the normalized time lapse between the whole train and the interrupted version. Thus -in order to have revivals- the time translation $y=\Delta t/\tau$ must be larger than the time lapse $x$.

Comparing two interrupted sequences with lapses respectively $x$ and $x'$, we generate two interferential returns corresponding respectively to $x$ and $x'$ and thus separated by $(x'-x)$.

The wave character, that in particle dynamics is associated with $k=p/h$, here is due to the duration of the sequence to be synchronized with the initial reference of duration $T$. Thus it seems bound to the theta –gamma cross modulation.

Let us now apply the above formalism to a linguistic task. A linguistic task consists of the comparison of two words, one corresponding to the last presentation, and the previous one recovered by the short memory within 2-3 s [13].

The words are coded as trains of neuronal spikes. Performance of the linguistic task amounts to synchronization of the two word trains. Take $T$ as the time duration of the second word. The previous one is interrupted at $\Delta T<T$ by the theta-gamma cross modulation. From what said above such a word spans a region of a functional space, that corresponds to a finite-dimensional Hilbert space.

The total spike train belongs to a Hilbert space of $2^T$ dimensions. It is represented by the ket $|T>$. The spike train interrupted after a duration $\Delta T$ provides a set of states living in the same $2^T$-dimensional Hilbert space. It is entangled with all possible realizations of *1*s and *0*s in the complementary interval $T-\Delta T$.

For sake of reasoning, let us consider the minimal interruption, that is $\Delta T=T-1$. In such a case, the synchronization task of $|T>$ with $|\Delta T>$ amounts to comparing $|T>$ with the entangled state

$$|E> = \frac{1}{\sqrt{2}}(|\Delta T, 0> + |\Delta T, 1>) \qquad (12)$$

and then performing a measurement based quantum computation [14, 15].

In general, $T-\Delta T=N$, hence synchronization amounts to measuring over a cluster of $2^N$ entangled states, that is, comparing the whole train of duration $T$ with $2^N$ different interrupted versions of it each one displaying differences from the original $T$ train and hence defeating full synchronization.

Thus, we model a word recognition process as the comparison between a reference word living in a finite dimension Hilbert space of $2^T$ dimensions, and a tentative word retrieved via the short term memory and interrupted to $\Delta T$ by the theta-gamma EEG cross modulation.

If we compare this process with the general approach of measurement-based quantum computation , there, once a cluster (h) has been prepared, its component states must be coupled by some interaction. As a fact, the theta-gamma modulation creates the cluster (h) of entangled states; but the successive synchronization *S* selects state |*d*) and thus it would be an irrelevant operation. We postulate that –before *S* is applied- the entangled states are coupled by emotional operators *€* [16,17]

Thus our SFSS operated language guess consists of the following sequence:

i) the interruption $\Delta T$ yields $2^{(T-\Delta T)}$ entangled states |h>;

ii) each one of these states is modified by the emotional coupling *(€)*:

$$€ |h> \rightarrow |h^*> \qquad (13)$$

iii) the synchronization S selects the state |h*> that best synchronizes to |d> (congruence). Without ii), the choice of |h> due to |d> would be a trivial operation.

As discussed above, quantizing the spike train implies a time interruption. As a fact, spikes occur at average rates corresponding to the EEG gamma band (around 50 Hz). However, superposed to the gamma band, a lower frequency background (theta band, around 7 Hz) controls the number of gamma band bursts. For instance, gamma power in the hippocampus is modulated by the phase of theta oscillations during working memory retention, and the strength of this cross-frequency coupling predicts individual working memory performance [8].

We here hypothesize that emotional effects raised by the first piece of a linguistic text induce a theta band interruption of the gamma band bursts, thus introducing an entanglement feature that speeds up the exploration of the semantic space in search of the meanings that best mutually match. In this behaviour, emotions do not have an aesthetic value "per sé", as instead maintained by neuro-aesthetic approaches [18], but rather they introduce the feature necessary to provide a fast scanning of all possible meanings within a de-coherence time. Hence the final decision does not depend on the emotions raised by the single word, but it is the result of the comparison of two successive pieces of a linguistic sequence (congruence). Such a behaviour, yielding the best match between the meanings of different words, was called "inverse Bayes procedure" [3].

We summarize our model of a linguistic endeavour. A previous piece of a text *(h)* is retrieved by the short term memory [13], modified by emotions *(€ )* into *(h⋆)* and compared via synchronization (S) with the next piece *(d).* The most adequate fit *P(d|h)* emerges as a result of the comparison (judgment and consequent decision).While in perceptual processes based on Bayes inference the conditional probability *P(d|h)* is stored in the long term memory, in language operations *P(d|h)* results from the comparison between *d* and *h\** (inverse Bayes procedure).

We stress the revolution brought about by the linguistic processes in human cognitive endeavours, namely; the SFSS is a peculiar process that cannot be grasped in terms of position-momentum variables; hence, the quantum constant for spike train position-duration uncertainty has nothing to do with Planck's constant.

The minimal energy disturbance which rules the de-coherence time is by no means $k_B T_R$ ($T_R$ being the room temperature); rather, since it corresponds to $\Delta P=1$, it entails the minimal energy necessary to add or destroy a cortical spike. This energy corresponds to the opening along the axon of about $10^7$ ionic channels each one requiring an ATP $\rightarrow$ ADP+P chemical reaction involving 0.3 eV [10], thus the minimal energy disturbance in neural spike dynamics is around $10^8 k_B T_R$. This is the evolutionary advantage of a brain, that is, to live comfortably at room temperature $T_R$ and yet be barely disturbed, as it were cooled at $10^{-8} T_R$.[19].

The procedure here described for connecting two successive pieces of a linguistic text explains other reported evidences of quantum effects in human cognitive processes, so far lacking a plausible framework. Models of quantum behaviour in language and decision taking have already been considered, starting 1995[20-22]. Most references are collected in a recent book[23]. The speculations introduced to justify a quantum speed up can be grouped in two categories ,

i) either they lack a dynamical basis and thus do not discuss limitations due to a quantum constant, hence, they do not inquire for a de-coherence time terminating the quantum operation;

or

ii) they refer to the quantum behaviour of Newtonian particles[24, 25] and hence are limited by a de-coherence time estimated around $10^{-14}$ s, well below the infra-sec scale of the cognitive processes.

In conclusion, while in the perceptual case, the cognitive action combines a bottom-up signal provided by the sensorial organs with a top-down interpretation provided by long term memories stored in extra-cortical areas, in the linguistic case, the comparison occurs between the code of the second piece and the code of the previous one retrieved by the short term memory. In this case , theta –gamma cross modulation introduces an information-time uncertainty (**ITU**), hence an entanglement among different words that provides a fast quantum search of meanings. In SFSS, a quantum behaviour entails pairs of incompatible observables, inducing measurement uncertainties bound by a novel constant. The emerging de-coherence time is compatible with the observed processing times in linguistic endeavours. On the contrary, all previously reported approaches either overlook the need for a quantization constant , or they quantize Newtonian particles by using Planck constant and consequently arrive to very short de-coherence times

## Example of a linguistic process

Let the word of the second piece be

$$|d>=|T> \qquad (14)$$

(living in a Hilbert space H* of $2^T$ dimensions)

and the word of the first piece be interrupted at ΔT=T-2, that is,

$$|h> = |\Delta T>  \quad (15)'$$

( living in a Hilbert space H** of $2^{\Delta T}$ dimensions).

Report the first word to T bits

$$|h> = |\Delta T> + ¼(|00>+|01>+|10>+|11>) \quad (16)$$

Now, if the last two bits of |d> are |10>, then, synchronization S of |h> with \d> leads toe table 1

*Table 1-trivial synchronization*

| States in T-ΔT | synchronization <h|d> |
|---|---|
| 00 | 0 |
| 01 | 0 |
| 10 | 1 |
| 11 | 0.5 |

As a result, we have maximal congruence for |h>=|ΔT> +|10>, that coincides with |d> ;this result is trivial. If however before synchronization we apply the operator emotional coupling € with the following weights (Table 2)

*Table 2-Role of the emotional operator €*

| States in T-ΔT | weight € |
|---|---|
| 00 | 1/4 |
| 01 | 1/4 |
| 10 | 0 |
| 11 | 1/2 |

--------------------------------------------------------------------------------

Then, the combined effect of € and S consists of recovering a state

$$|h'> = |\Delta T > + |11> \quad (17)$$

Summarizing, the sequence of operations

i)   ΔT (interruption);
ii)  € (different weights for complementary states in space H*-H**, whence |h>→\h'>.
iii) **S** (synchronization <d|h'>)

yields the novelty

$$|h'\rangle \neq |h\rangle \qquad (18)$$

## Supplementary material - How SFSS operates in brain processes

We outline the dynamical processes occurring in single neurons and how neurons couple to provide a coherent picture.

i) Within a neuron, voltage travels via an axon that operates as a nonlinear delay line; precisely the voltage travels as a soliton spike with the energy loss replaced by extra energy provided by the opening along the axon of about $10^7$ ionic channels, each one entailing an ATP →ADP+P chemical reaction that releases 0.3 eV [10].

ii) In single neurons, homoclinic chaos yields identical spikes non-uniformly distributed in time [27]; the inter-play between gamma and theta bands of EEG, contribute to tailoring the sequence length [8, 9] ; neurons get into speaking terms by synchronizing their spike sequences over a finite time interval (whence SFSS= synchronization (of) finite spike sequences) [28, 29].

iii) The role of SFSS in brain computation; specifically, the role of a synchronization reader (called GWS= global workspace [30,31]).

iv) Difference between apprehension, modelled as a Bayes inference, and a linguistic task, modelled as an inverse Bayes inference [3,4].

Let us explore in detail the processes just outlined.

The single neuron dynamics can be modelled [27] as a homoclinic orbit in a 3D phase-space, returning to a saddle point S with an approach rate $\alpha$ and an escape rate $\gamma$. If $\gamma > \alpha$, then the orbital period is chaotic, otherwise it is regular. The 3D orbit projected onto a single axis yields a standard spike voltage (neural case: 100 mV, 1 ms) that repeats in time with a minimal separation (bin) of 3 ms. As a fact, the phase-space position of the saddle S , that drives the escape time from the saddle and hence the inter-spike separation, is affected by the local voltage provided by the neighbouring cortical areas. In the gamma band (from 40 to 80 Hz), spike trains are clustered around each peak (where the S position is conveniently high) and then absent up to the next peak; this yields clusters of spikes separated by wide empty intervals. The gamma band peaks are further modulated by a lower frequency theta band that pushes down some of the gamma maxima, thus introducing new gaps in the spike sequence [8, 9].

Synchronization of the spike sequence over a finite time (SFSS) is the way different brain areas exchange information [1,4]; synchronized cortical areas represent a suitable response to external stimuli (feature binding). Synchronization of a single cortical area results from the interplay between sensorial stimuli and the local potential provided by the combination of gamma and theta EEG bands.

In case of competition between two different cortical areas providing different amounts of synchronization, a Global Work Space (GWS) reads the corresponding spike sequences and decides the more appropriate one to trigger further actions [30 31]

One must distinguish two separate moments of human cognition, namely, *apprehension* (*A*) whereby a coherent perception emerges from the recruitment of neuronal groups stimulated by sensorial stimuli, and *judgment (B)* whereby memory recalls previous (*A*) units coded in a suitable language and from comparison of two coded *A* units it follows the emergence of meaning and hence a judgment.

The first moment *(A),* has a duration around 1 s [5]; its associated neuronal correlate consists of the synchronization of the EEG signals in the gamma band coming from distant cortical areas [6]. It can be described as an interpretation of the sensorial stimuli on the basis of available algorithms, through a Bayes inference[32]. Precisely, calling *h*(*h*= hypothesis) the interpretative hypotheses in presence of a sensorial stimulus *d* (*d*=datum), the Bayes inference selects the most plausible hypothesis *h\**, that determines the motor reaction, exploiting a memorized algorithm *P(d|h)* representing the conditional probability that a datum *d* be the consequence of an hypothesis *h* [33].

The *P(d|h)* is the equipment whereby a cognitive agent faces the world. By equipping a robot with a convenient set of *P(d|h),* we expect a similar behaviour.

The second moment *(B)* entails a comparison between two apprehensions (*A*) acquired at different times, coded in a given language and recalled by the memory. If, in analogy with (*A*), we call *d* the code of the second apprehension and *h\** the code of the first one, now- at variance with (*A*)- *h\** is already given; instead, the relation *P(d|h)* which connects them must be retrieved; it represents the *conformity* between *d* and *h\*,* that is, the best interpretation of *d* in the light of *h\*.*Thus, in linguistic operations, we compare two successive pieces of the text and extract the conformity ( that is, *the meaning*) of the second one on the basis of the first one.

We have recalled that trains of identical spikes positioned at unequal times represent the electrical activity of a cortical neuron. Thus, a neuronal signal is a binary sequence of *0*'s and *1*'s, depending on whether a given bin is empty or filled by a spike. Each cortical neuron has two ways to modify its spike occurrence, either coupling to other cortical neurons or receiving signals from extra-cortical regions. Thus, a meaningful linguistic piece is coded by a train of neural spikes.

In the perceptual case (*A*), a relevant conjecture, called "feature binding"[27, 28], provides a sound guess on how the spike trains in distant neuronal areas get synchronized.

Any linguistic item consists of successive pieces to be compared. Precisely, a short term memory mechanism recalls the previous piece and compares it with the next one. Comparison consists of a synchronization process between the trains. If a word has *N* different meanings in our private memory, the most appropriate meaning is the one that has the largest synchronization with the next piece.

In virtue of the theta-gamma modulation, the spike train coding the first word is interrupted from a duration *T* to a duration Δ*T*<*T*. To perform the synchronization, it must be lengthened by *T*−Δ*T* and this can occur in *N*=2$^{(T-\Delta T)}$ ways by filling the *T*−Δ*T* interval with *N* different sequences of *0* and *1*. Thus the first word is coded as a cluster state |*E*> consisting of *N* different sequences. As explained above, state |*E*> is entangled, hence, the *N* component states have quantum correlations.

Cognitive dynamics, *SPIE Noise and Fluctuation in Biological, Biophysical, and Biomedical Systems* – Paper 6602-36 (2007).

## Acknowledgments

I am indebted to Augusto Smerzi and Alessandro Farini for very helpful discussions.